\documentclass[fleqn]{aa}
\usepackage{graphicx,natbib}
\usepackage{amsmath,amssymb}
\bibpunct{(}{)}{;}{a}{}{,}
\begin{document}
\title{Near-IR coronagraphic imaging of the companion to HR\,7672\thanks{Based on data collected at the Palomar 200-inch}}
\author{A. Boccaletti\inst{1}, G. Chauvin\inst{2}, A.-M. Lagrange\inst{2} and F. Marchis\inst{3}}
\institute{LESIA, Observatoire de Paris-Meudon, 5 pl J. Janssen, F-92195 Meudon, France\\
\email{Anthony.Boccaletti@obspm.fr}
 \and Laboratoire
d'Astrophysique de l'Observatoire de Grenoble, BP53 F-38041
Grenoble Cedex 9, France \\
\email{gael.chauvin@obs.ujf-grenoble.fr},
\email{anne-marie.lagrange@obs.ujf-grenoble.fr} \and University of
California at Berkeley, Department of Astronomy, 601 Campbell
Hall, Berkeley, CA 94720, USA \\
\email{fmarchis@astron.berkeley.edu}}
\titlerunning{Coronagraphic imaging of HR\,7672}
\authorrunning{Boccaletti A., et al.}
\offprints{A. Boccaletti}
%%%%%%%%%%%%%%%%%%%%%%%%%%%%%%%%%%%%%%%%%%%%%%%%%%%%%%%%%%%%%%%%%%
\abstract{This article presents coronagraphic images of the
low-mass companion to the star HR\,7672 observed at the Palomar
200 inch telescope and first detected at Gemini and Keck in the K
band by \citet{liu02}. We obtained additional photometry in
$J$($1.2\mu m$), $H$($1.6\mu m$) and $K_s$ ($2.2\mu m$) bands to
cover the full near-IR domain and hence to further constrain the
nature of the companion. A mass estimate of 58-71$\ \mathrm{M_J}$
is derived from evolutionary models of very low-mass objects.

} \maketitle
%%%%%%%%%%%%%%%%%%%%%%%%%%%%%%%%%%%%%%%%%%%%%%%%%%%%%%%%%%%%%%%%%%
\section{Introduction}
Since the discovery of the first bona fide substellar objects:
GD165\,B \citep{Becklin88} and Gl229\,B \citep{Nakajima95}, an
ever growing number of low-mass objects are routinely detected.
Large IR surveys like 2MASS \citep{Skrutskie97}, DENIS
\citep{Epchtein97} and SLOAN \citep{York00} or dedicated surveys
in star-forming regions were initiated to explore the bottom of
the main-sequence and have provided a consistent sample of more
than 200 isolated low-mass stars. In the Trapezium young cluster,
the presence of planetary mass candidates has been announced
independently by \citet{Lucas00} and \citet{Zapatero00} but
remains doubtful since the distances of these candidates are
uncertain \citep{Tinney03}.

Two distinct classes of stars later than M have been identified:
the so-called L types and T types. The L class, proposed by
\citet{Kirkpatrick99}, gathers objects with effective temperatures
 lower than M dwarfs ($T_\mathrm{eff}<2000\ \mathrm{K}$) and
 featuring strong lines of neutral alkali
elements, absence of TiO and VO absorption and stronger H$_2$O
absorption in the visible. As a consequence, the near-IR colors
are becoming very red with $0.5<H-K_s<1$ \citep{Kirkpatrick00}.
According to their age, some L dwarfs can also be Brown Dwarfs
(BDs) since their mass is often below the stellar boundary (0.075
$M_{\sun}$). Independently of the definition of the L class given
by \citet{Kirkpatrick99}, \citet{Martin99} have proposed a
different classification using a temperature scale derived by
\citet{basri00}. The two approaches yield different
classifications for later types (beyond L2) and this discrepancy
demonstrates the difficulty to tackle this new subject in
astrophysics. \citet{Kirkpatrick99} also proposed to designate
Gl229\,B-like objects as T dwarfs. These objects, also called
"methane dwarfs", are cooler than the L type
($T_\mathrm{eff}<1300\ \mathrm{K}$) and exhibit very strong H$_2$0
and CH$_4$ absorption in the $H$ and $K$ bands. As a result, their
color index $J-K$ and $H-K$ are very close to 0. It is now well
admitted that these 2 classes of stars correspond to the missing
part at the bottom of the main sequence between M stars and giant
planets. The study of low-mass objects either isolated or orbiting
stars is of course very exciting and is definitely mandatory to
further understand both the brown dwarf and planetary formation
process.

%% les modeles: burrows, chabrier, allard
Low-mass stars as well as brown dwarfs have also been studied from
a theoretical point of view. However, the processes occurring in
these objects appear quite complex (especially regarding the
presence of dust below $T_\mathrm{eff}<2800\ \mathrm{K}$) and
require new type of models. Non grey atmosphere models have been
developed by \citet{burrows97} and \citet{Chabrier97}. Dust-free
atmosphere models agree well with stellar objects down to M stars,
but the dust needs to be taken into account for lower masses. A
productive group \citep{baraffe98,baraffe02,chabrier00} has
calculated evolutionary models including dust in the atmosphere in
order to provide the magnitudes of sub-stellar objects at
different ages that could be compared directly to observations.
The treatment of dust is somewhat difficult and different models
are not clearly distinguishable when compared to data points. In
addition, the optical part of the T dwarfs spectrum is
complementary to the near-IR regarding spectral typing and models
were proposed by \citet{burrows02}.
\begin{figure*}[th!]
\label{imagescoro} \centering
\includegraphics[width=5cm]{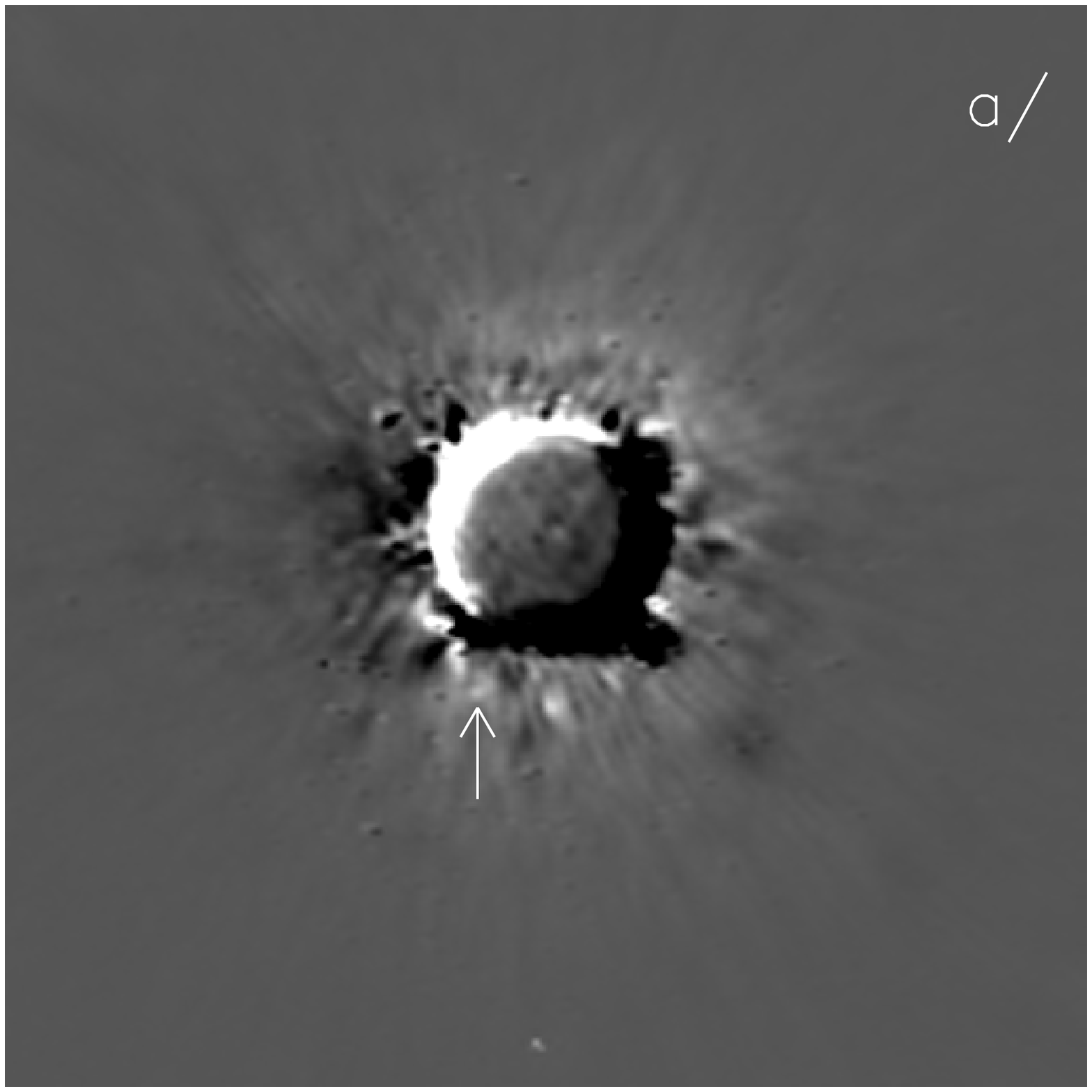}
\includegraphics[width=5cm]{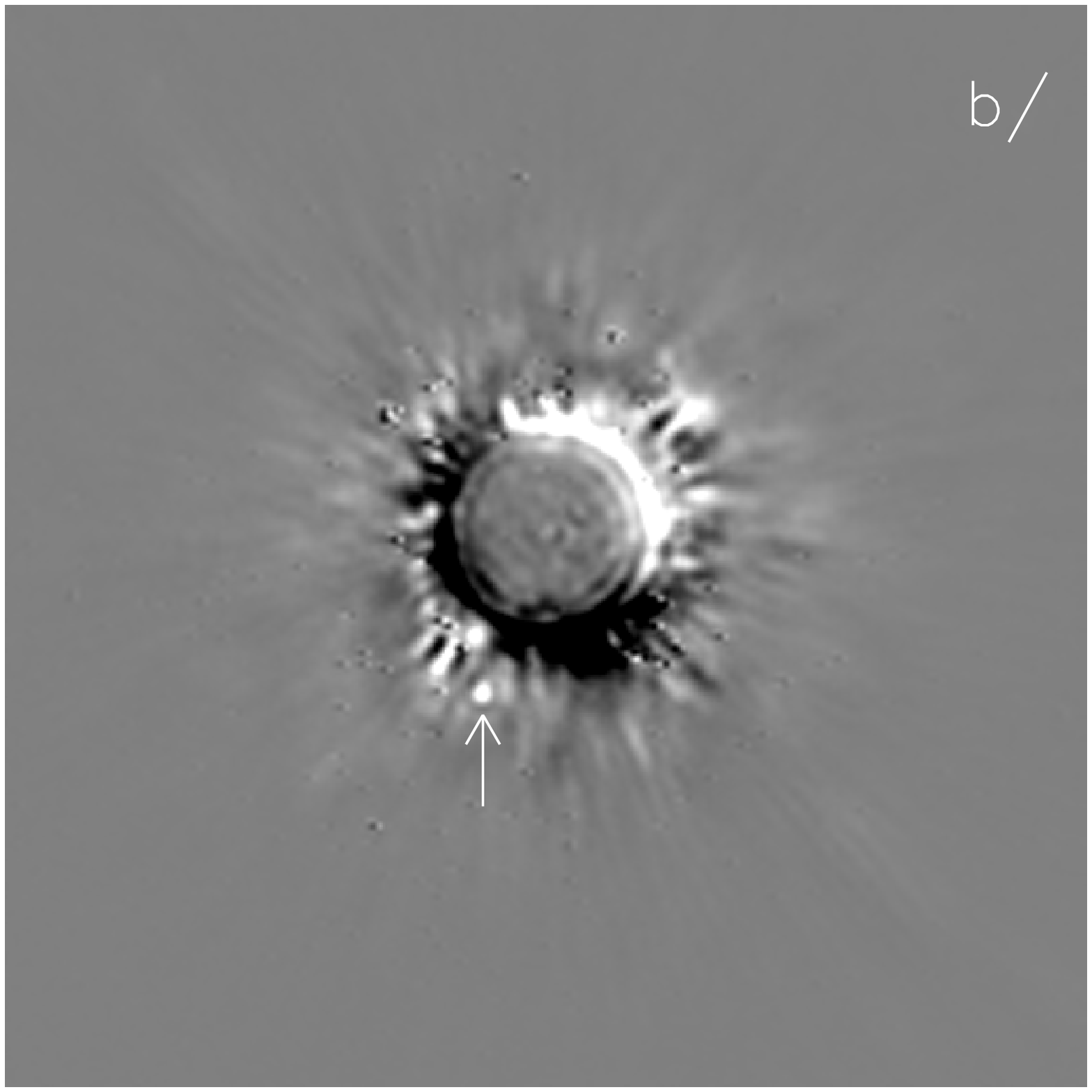}
\includegraphics[width=5cm]{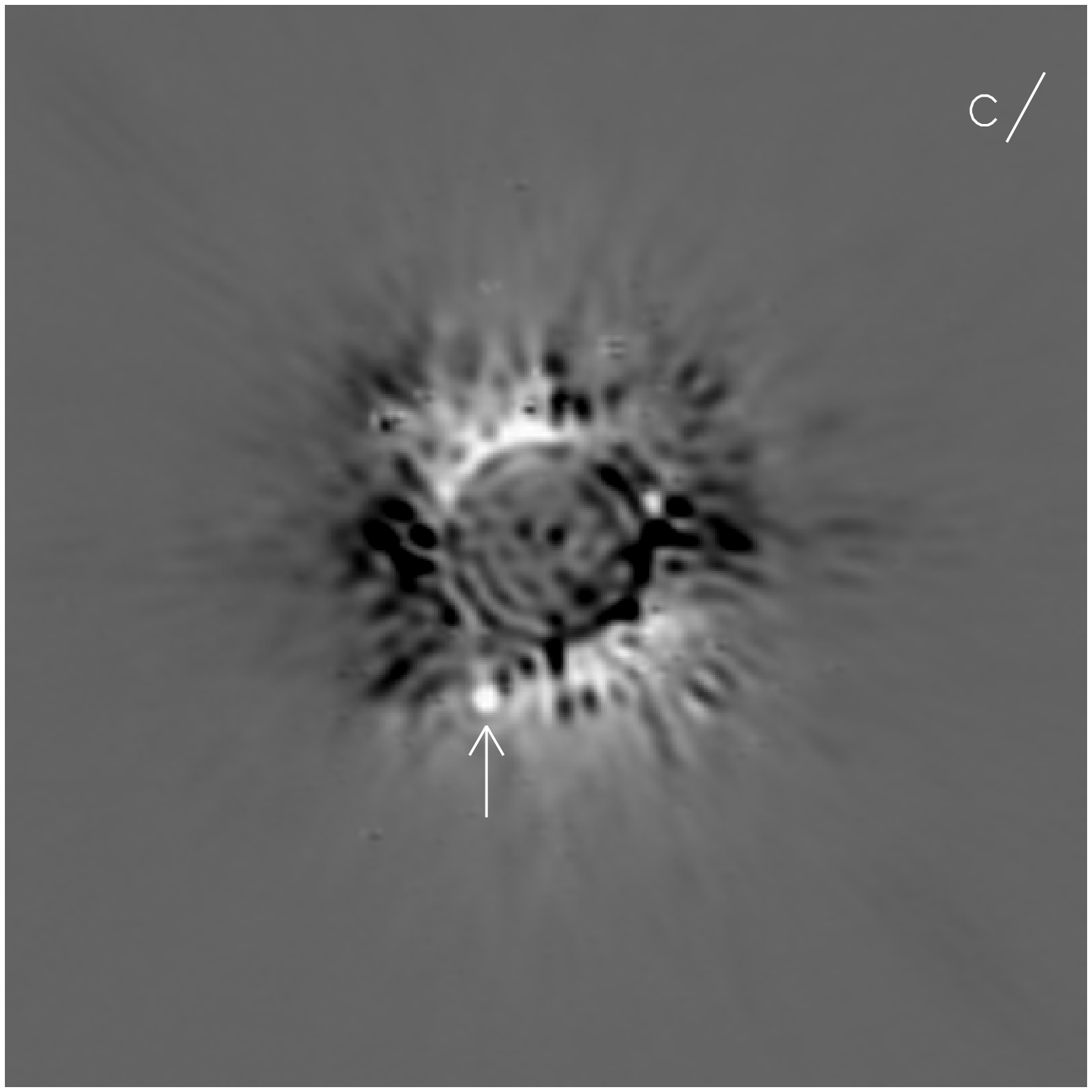}
\caption[]{Coronagraphic images of HR7672 subtracted with the
calibration star HD\,194012 for respectively $J$ (a/), $H$ (b/)
and $K_s$ (c/) filters. The field of view is $5" \times 5"$. North
is up, East is left.}
\end{figure*}

%% BD desert ....
Radial velocity surveys have evidenced the so-called "Brown Dwarf
desert" at small separations $<4\ \mathrm{AU}$
\citep{halbwachs2000}. At very wide separations larger than $1000\
\mathrm{AU}$, a few BD companions were discovered
\citep{Kirkpatrick01} by the 2MASS Survey suggesting that this
brown dwarf desert may not exist \citep{gizis01}. Intermediate
separations can be probed by AO direct imaging as demonstrated by
the case of Gl\,86B, a possible BD orbiting a planetary system at
$18.75\ \mathrm{AU}$ \citep{els01}.
 The problem is to know whether or not this desert is caused by a
 detection bias or is a direct consequence of
BD formation. The gravitational influence of a BD close to the
star may markedly impact on the history of a circumstellar disk
and hence on the planetary formation \citep{arty98}. Increasing
the sample of bounded BDs is therefore crucial for the
understanding of substellar formation. Although substantial
progress are still needed, high-contrast technics are efficient
tools to address these basic questions.

%% rappel decouverte de HR7672b par Liu et al. 2002
In 2002, Liu et al. have reported the discovery of a cool low mass
object as a companion candidate to the star HR\,7672 using
Adaptive Optics (AO) imaging. Two epochs were obtained in 2001 at
Gemini-North and KeckII telescopes and the proper motion of the
companion was found correlated with that of the star. Images and
spectra were obtained in the $K$ band from which they derived a
spectral type L$4.5\pm1.5$. Using the theoretical models of
\citet{burrows97} and \citet{chabrier00} they determined a mass of
$55-78\ \mathrm{M_J}$.

%% descriptif du papier
We obtained multi-wavelength high-contrast images of the companion
to HR\,7672 in order to confirm and further constrain its spectral
type and mass. The results are presented in this paper. Section
\ref{observations} describes the coronagraphic observations and
photometric measurements are given in section \ref{photometry}.
Finally, a mass estimate is given in section \ref{mass}.
%%%%%%%%%%%%%%%%%%%%%%%%%%%%%%%%%%%%%%%%%%%%%%%%%%%%%%%%%%%%%%%%%%
\section{Observations and Standard Reductions}
\label{observations} The nearby star HR\,7672 (G0\,V, $V=5.80$,
\citet{Gray01}) was observed on July 17, 2002 at the Palomar 200
inch telescope. The data were obtained with PALAO the
241-actuators AO system \citep{troy00} installed at the Cassegrain
focus and PHARO the near-IR camera \citep{hayward01}. A Lyot
coronagraph including 2 opaque masks ($\varnothing$ 0.91" and
0.41") with corresponding stops is located inside the cryostat of
the camera and is dedicated to high-contrast imaging. The detector
has a pixel sampling of $0.025$ mas.

First of all, the star was observed with the $J$, $H$ and $K_s$
filters without the coronagraph to provide a photometric
calibration. In that case, a neutral density was necessary to
avoid saturation. The Strehl ratio delivered by the AO system on
that night was about 15\%, 32\% and 44\% with FWHM of 51, 61 and
82 $mas$ for respectively $J$, $H$ and $K_s$ bands. Coronagraphic
images of HR\,7672 were then obtained with the $0.91"$ mask. Both
coronagraphic and non-coronagraphic images were obtained with the
same pupil stop (the so-called medium cross) to allow quantitative
comparison. A calibration star (HD\,194012, F8\,V, $V=6.17$) was
also observed for each filter to allow subtraction of the residual
static speckles. This calibration star is chosen angularly close
and of similar magnitude and spectral type to ensure identical
compensation by the AO system. The subtraction process is very
efficient to remove the residual diffraction and the quasi-static
speckles but critically depends on the precise centering of the
star behind the coronagraphic mask. Images of the target and the
calibration star were then corrected from the bad pixels, the flat
field and were subtracted with a median sky frame. To properly
perform the subtraction process, the calibration star was first
recentered and then scaled to the intensity of the target. This
scaling factor is derived from the coronagraphic images. The
coronagraphic images of HR\,7672 reduced with the above procedure
are displayed on Fig. 1. The angular separation and position angle
of the companion derived from the $JHK_s$ images are $\rho=788\pm
6$ mas and $PA=156.6\pm0.9^\circ$ which is consistent with the
astrometric measurements of \citet{liu02}. Using the coronagraphic
images of Fig. 1, we compared the noise level in a ring of
$\lambda/D$ at each angular radius $\rho$ with the maximum
intensity of the star (no mask) to provide the companion
detectability displayed of Fig. 2.

\begin{figure}[t]
\label{seuil} \centering
\includegraphics[width=8cm]{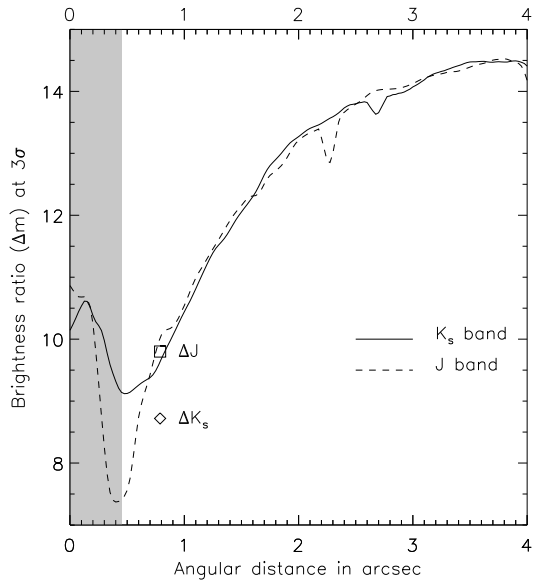}
\caption[]{Detectability of a companion in the field around the
primary star. Each curve gives the contrast at $3\sigma$ in
difference of magnitudes ($J$ and $K_s$) and as a function of the
angular radius. The magnitude differences ($\Delta J$, $\Delta
K_s$) of the companion to HR\,7672 are overplotted.}
\end{figure}

%%%%%%%%%%%%%%%%%%%%%%%%%%%%%%%%%%%%%%%%%%%%%%%%%%%%%%%%%%%%%%%%%%
\section{Near-IR photometry of HR\,7672\,B}

\label{photometry} First of all, we measured the flux of the
primary star on non-coronagraphic images. We used Gl\,777\,A, a
star observed the same night, as a photometric calibrator. Its
near IR magnitudes according to the Catalogue of IR Observations
\citep{Gezari99} are $J=4.45$, $H=4.11$ and $K=4.05$ (with
$\sim0.02~mag$ uncertainty). The fluxes of the stellar PSFs are
calculated in a circular integrating window with a size ranging
from $4.4\ \mathrm{\lambda/D}$ (16 pixels) to $8.8\
\mathrm{\lambda/D}$ (32 pixels) thus providing a photometric
uncertainty of about 0.01 magnitude. Then, the corresponding
photometry for HR\,7672 is $J=4.59\pm0.04$, $H=4.40\pm0.02$ and
$K_s=4.32\pm0.02$. For the $J$ band no PSF of Gl\,777\,A was
available. So, we derived the photometry of HR\,7672\,A using the
coronagraphic data alone leading to a larger uncertainty. That
probably explains the discrepancy of the color index $J-H=0.19\pm
0.04$ with that of a standard star ($J-H=0.305$,
\citet{Bessell88}).

To compute the intensity of HR\,7672\,B we compared the flux of
the primary on direct images with that of the companion detected
on the coronagraphic images. The integrating window varies from
$2.44\ \mathrm{\lambda/D}$ to $4.88\ \mathrm{\lambda/D}$ to derive
the error bars of the HR\,7672\,B flux ratio. Such a small
integrating window is necessary to avoid the pollution originating
from coronagraphic residuals of the primary star. The measurement
in the $J$ band is more critical since the peak of the companion
has about the same brightness than speckle residuals from the
primary and its detection was actually made possible owing to the
$H$ and $K_s$ images. In that case the integrating window was
narrowed down to 1-5 pixels in diameter. However, we are still
expecting the flux measurement in the $J$ band to be
overestimated. The results are summarized in Tab. 1. We obtained a
brightness ratio of $\Delta J=9.80\pm 0.20$, $\Delta H=9.64\pm
0.14$ and $\Delta K_s=8.72\pm 0.10$. The photometric uncertainty
is of course much larger than for the primary star owing to the
residual flux in the coronagraphic image. The corresponding color
indexes are: $J-K_s=1.35\pm 0.22$, $H-K_s=1.00\pm 0.17$ and
$J-H=0.35\pm 0.24$.

\begin{table}[t!]
\label{table_deltaMb}
\begin{center}
\begin{tabular}{|l|c|c|c|} \hline
                  & $J(\ast)$              & $H$             & $K_s$     \\\hline
$\Delta m$        & $9.80\pm 0.20$   & $9.64\pm 0.14$  & $8.72\pm
0.10$\\\hline $m$   & $14.39\pm 0.20$  & $14.04\pm 0.14$ &
$13.04\pm 0.10$\\\hline $M$ & $13.16 \pm 0.20$ & $12.81\pm 0.14$ &
$11.81\pm 0.10$
\\\hline
\end{tabular}
\end{center}
\caption{Brightness ratio ($\Delta m$), visual magnitude ($m$) and
absolute magnitude ($M$) of the companion to HR\,7672.  ($\ast$)
$J$ band photometry appears to be corrupted by the huge speckle
background and is not used in this paper to derive spectral type
and mass of the companion.}
\end{table}

We used the data from \citet{Dahn02} to derive a linear
relationship between the absolute $JHK_s$ magnitudes and the
spectral type of an M and L dwarfs sample. These relations are
expected to be more accurate than the ones provided by
\citet{Kirkpatrick00} since they were using a smaller sample and
preliminary astrometry.
 The fitting of this sample (40 M and L stars, T stars removed)
leads to the following relations:
\begin{equation}
\label{relation_J} M_J= 8.472 + 0.327 \times Sp
\end{equation}
\begin{equation}
\label{relation_H} M_H= 8.212 + 0.279 \times Sp
\end{equation}
\begin{equation}
\label{relation_Ks} M_{K_s}=8.010  +0.247\times Sp
\end{equation}
where $Sp$ is 6.5 for M6.5\,V up to 18 for L8\,V. The fit
dispersion is about 1.2 subclass on X-axis and 0.33 magnitude on
Y-axis. This gives a spectral type of L3.5 to L5 for the $J$ band
(Eq. \ref{relation_J}), L6 to L7 for the $H$ band (Eq.
\ref{relation_H}) and L5 to L6 for the $K_s$ band (Eq.
\ref{relation_Ks}). If we consider that the $J$ band photometry is
corrupted by the speckle residue, the $H$ and $Ks$ data lead to
L6$\pm$1.5 including uncertainties and the fit scattering. Given
the error bars, this result is in agreement with L4.5$\pm$1.5
announced by \cite{liu02}.

%%%%%%%%%%%%%%%%%%%%%%%%%%%%%%%%%%%%%%%%%%%%%%%%%%%%%%%%%%%%%%%%%%
\section{Mass estimate}
\label{mass} A mass estimate can be obtained from photometric
measurements assuming an evolutionary model of low-mass stars.
\citet{baraffe98} and \citet{chabrier00} have developed 2 types of
models to describe the atmosphere of very low-mass stars. The
so-called COND and DUSTY models both account for the presence of
grains in the atmosphere. The DUSTY model also includes the grain
opacity and is more suited for objects with
$T_\mathrm{eff}\lesssim 2800\ \mathrm{K}$ while the COND model is
more appropriate to describe cooler methane objects
($T_\mathrm{eff}\lesssim 1300-1400\ \mathrm{K}$) as explained in
\citet{baraffe02}. Then, special care is needed since HR\,7672\,B
probably lies in between these temperature boundaries. Assuming an
age of 1 to 3 Gyr \citep{liu02} for both the primary and the
companion, we plotted the magnitude-color diagram displayed on
Fig. 3. The DUSTY model appears definitely more consistent with
the photometry of HR\,7672\,B at least in $H$ and $K_s$ bands.
However, the data obtained in the $J$ band are still conflicting
with the models for the same reason as above.
%%%%%%%%%%%%%%%
\begin{figure*}[t]
\label{CMD} \centering
\includegraphics[width=8.cm]{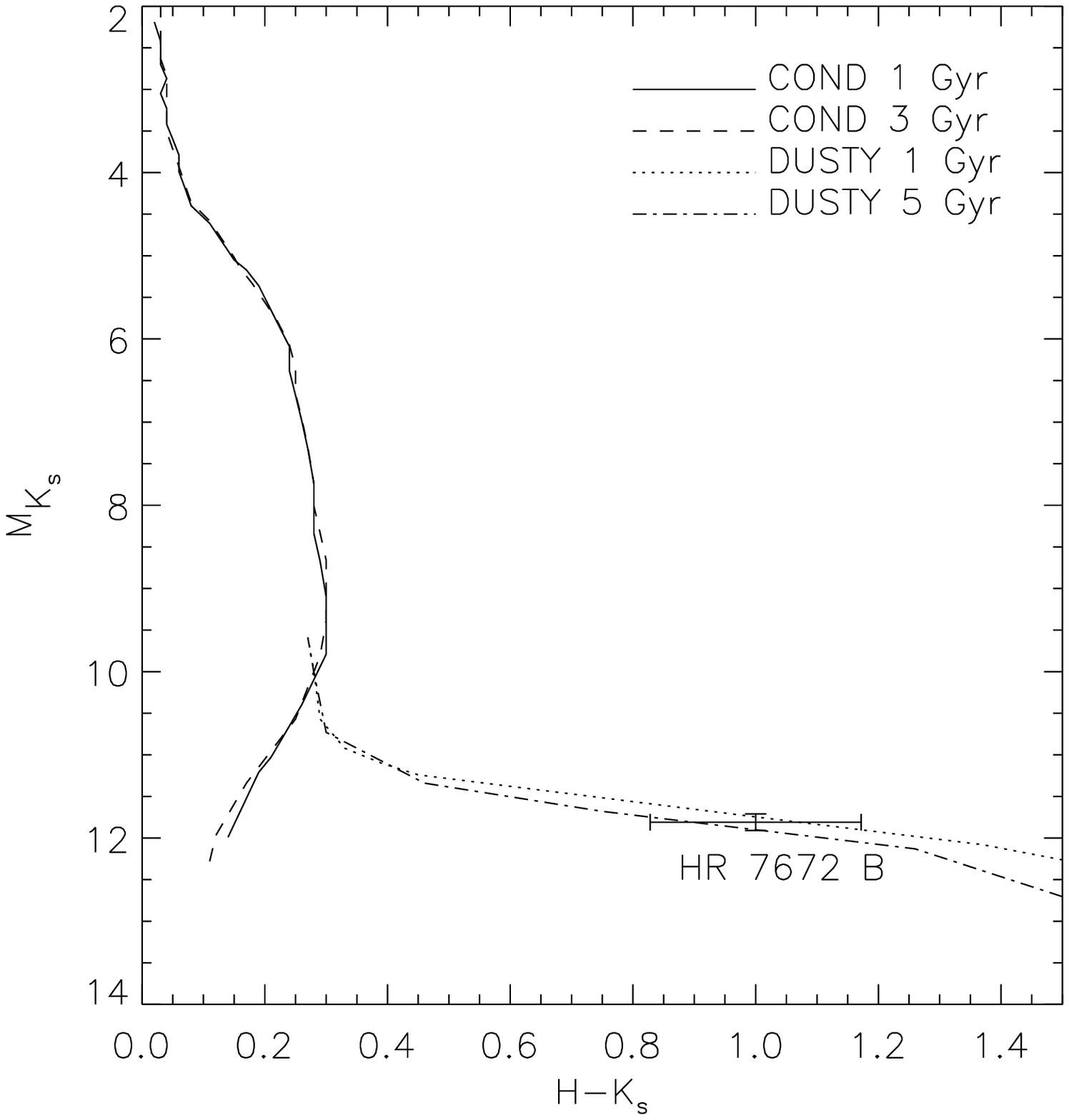}
\includegraphics[width=8.cm]{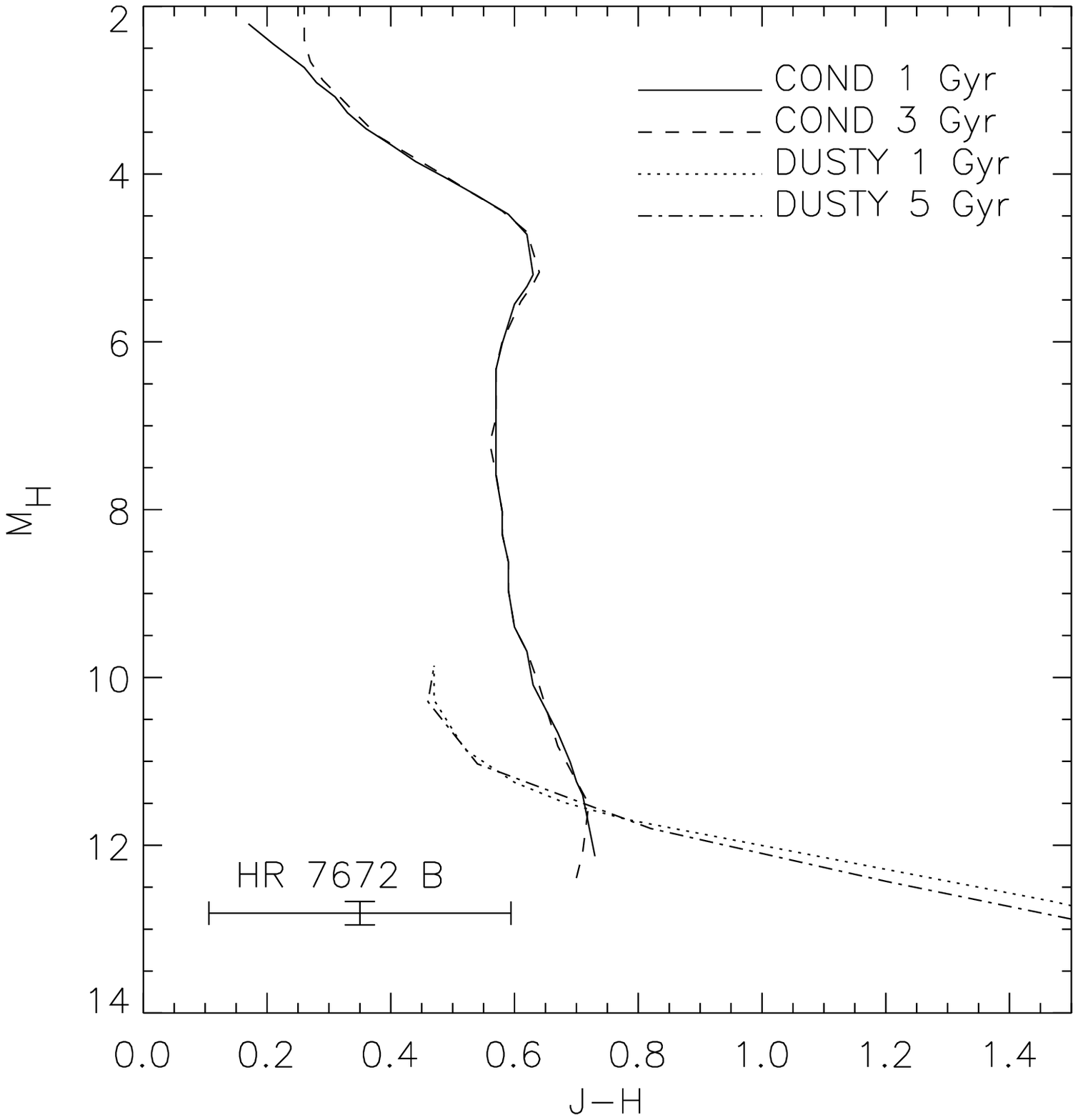}
\caption[]{Color-magnitude diagrams showing the difference between
the COND and the DUSTY models of \citet{chabrier00}. Based on the
$H$ and $K_s$ photometry we conclude that HR\,7672\,B was
definitely in better agreement with the DUSTY one.}
\end{figure*}
%%%%%%%%%%%%%%%%%%%%%%%%

We therefore used the isochrones of the DUSTY model
\citep{chabrier00} to plot the evolutionary diagrams in $J$, $H$
and $K_s$ bands for masses ranges between $50\ \mathrm{M_J}$ and
$80\ \mathrm{M_J}$ (Fig. 4,5). The $K_s$ and $H$ bands absolute
magnitudes yield consistent values ranging between 58 and 71$\
\mathrm{M_J}$. This slightly improves the result found by
\citet{liu02} using $K_s$ photometry alone (55-72$\
\mathrm{M_J}$), thus revising the mass towards larger values. The
$J$ band photometry is unfortunately not accurate enough and
obviously leads to higher masses between 64 and 73$\
\mathrm{M_J}$. However, although we are expecting an overestimated
flux in the $J$ band, the plot of Fig. 5 shows that in this range
of mass, the absolute magnitude increases very rapidly with the
age. Therefore, a large photometric error does not imply
systematically a large mass error. We can tentatively assess the
actual $J$ band photometry assuming that HR\,7672\,B follows the
same photometric relationships as field L dwarfs. Reversing Eq. 3,
the L6$\pm$1.5 spectral type should correspond to an absolute $J$
magnitude of 13.2 to 14.2. These new values are overplotted on
Fig. 5 and give a rough estimate of the mass : 60-72$\
\mathrm{M_J}$ in better agreement with the mass derived from $H$
and $K_s$.
%%%%%%%%%%%%%%%%%%%%%%%%%%%%%%%%%%%%%%%%%%%%%%%%%%%%%%%%%%%%%%%%%%
\section{Conclusions}
Despite a lower angular resolution and a lower sensitivity than
Gemini and the Keck, the high-order AO system of the Palomar 200
inch owing to the Lyot coronagraph has enable the detection of the
low-mass companion to HR\,7672 in the $J$, $H$ and $K_s$ bands.
This multi-wavelength photometric analysis was performed to
further constrain the characteristics of the companion with
respect to the previous work \citep{liu02}. By comparison with L
field stars we derive a spectral type of L6$\pm$1.5, although
similar accuracy was obtained by \citet{liu02} with both K-band
photometry and spectra. However, the 3-colors information is
required to check the consistency with the models. The
color-magnitude diagrams suggest that HR\,7672\,B is better
consistent with the DUSTY model of \citet{chabrier00} and hence,
should contain dust grains in the upper atmosphere. Then, using
the state-of-the-art evolutionary model we obtained a mass
estimate of 58-71$\ \mathrm{M_J}$ based on $J$, $H$ and $K_s$
photometry. This places HR\,7672\,B right below the
hydrogen-burning limit.
%The mass estimate is therefore larger by 3
%$\mathrm{M_J}$ than the one derived by \citet{liu02}.
Nevertheless, as we discussed in section \ref{photometry}, the $J$
band photometry may be questioned since the companion has an
intensity very similar to that of the local speckles. A non
negligible part of the stellar flux therefore contributes to the
flux measurement of the companion. Additional observations, with
NAOS \citep{lagrange03} on the VLT for instance, would definitely
provide a more accurate photometry at the shorter IR wavelengths
which is mandatory to limit the possible range of mass. However,
the mass derived from photometry is highly model-dependent. First,
the model should be compared with a large sample of field L stars
to check its consistency. Second, the case of the 2-body system
HR\,7672 is very interesting since it may allow to derive the
dynamical mass in a few years ($r=\mathrm{14\,AU}$). Coronagraphic
searches of faint companions to nearby stars are therefore very
important to improve the sample of binary objects.
%%%%%%%%%%%
\begin{figure*}[ht!]
\label{plotmass} \centering
\resizebox{8cm}{8cm}{\includegraphics[width=8cm]{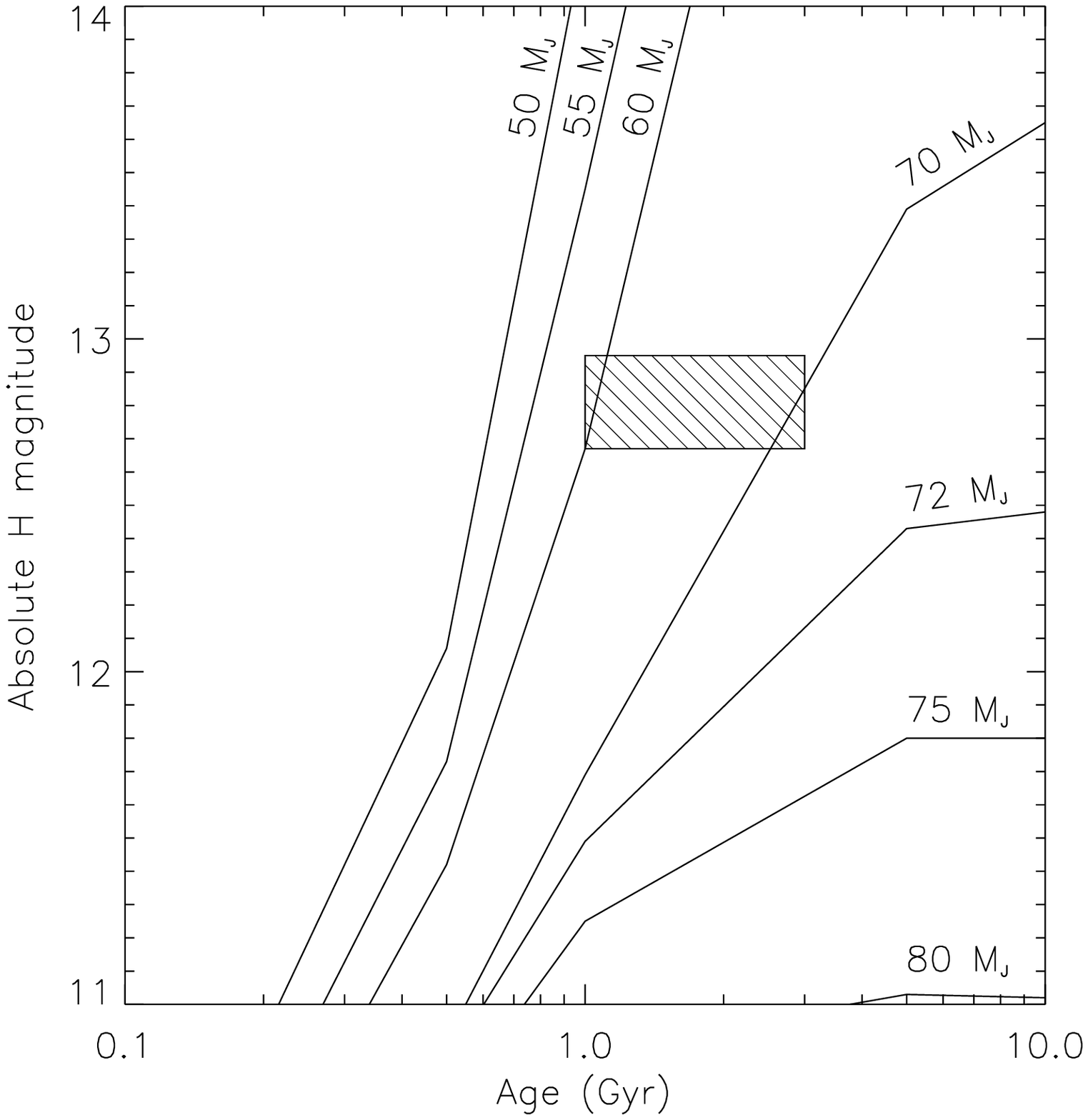}}
\resizebox{8cm}{8cm}{\includegraphics[width=8cm]{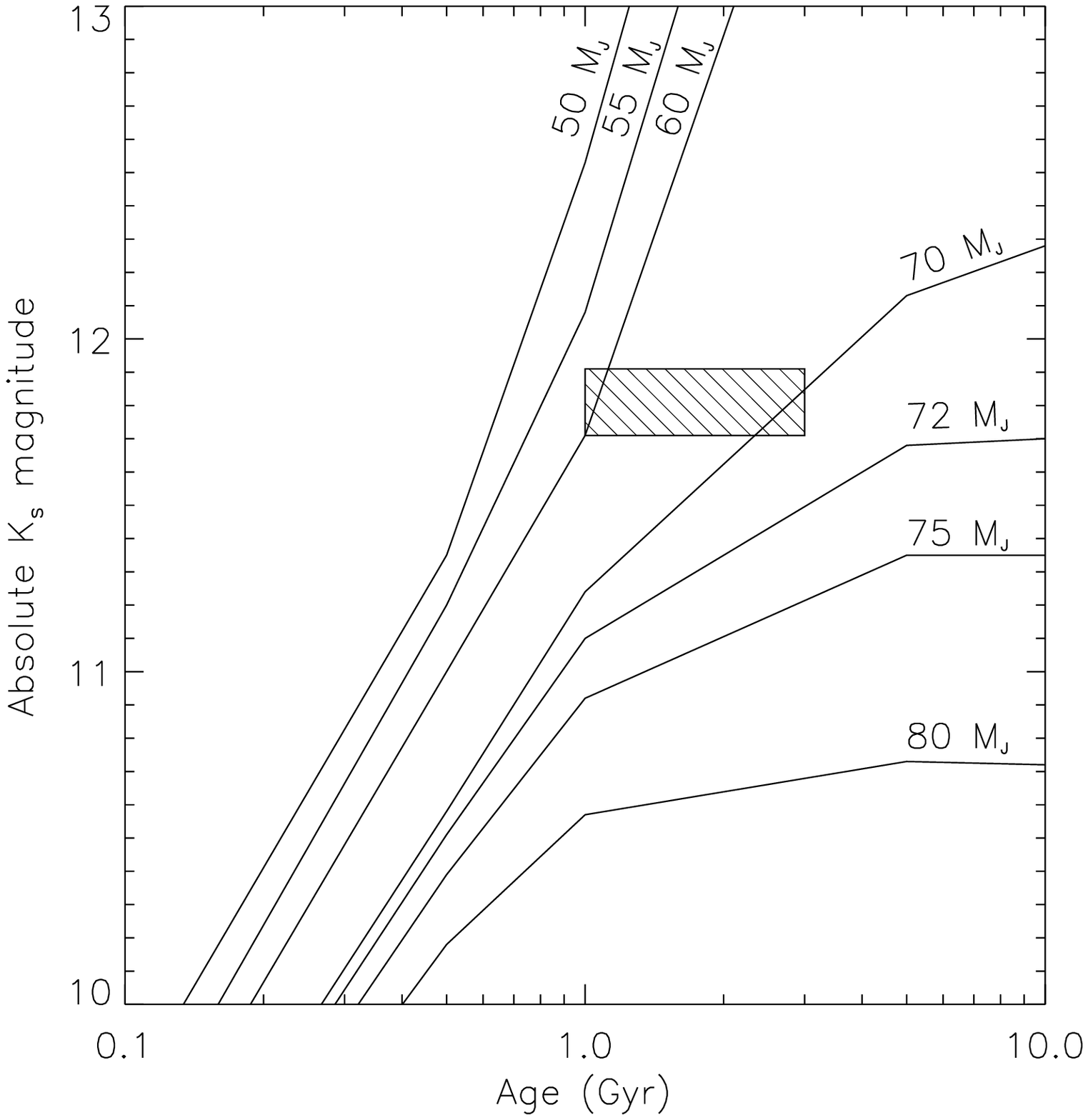}}
\caption[]{Absolute magnitude in $H$ (left) and $K_s$ (right)
filters as a function of the age (in Gyr) assuming the DUSTY model
of \citet{chabrier00}. The error box for HR\,7672\,B is
overplotted assuming an age of 1 to 3 Gyr.}
\end{figure*}
%%%%%%%%%%%%
If L and T dwarfs have been intensively observed in the near IR,
very few data are available on their visible spectrum except in
some particular cases (Gl\,229\,B for instance). \citet{Dahn02}
have carried out a photometric study in the visible and near-IR of
L and T dwarfs with known distances and found that the combination
of visible and IR (especially $I-J$) is more accurate than $JHK_s$
alone to derive the absolute magnitude of late M and L dwarfs.
Therefore, visible observations of HR\,7672\,B would be very
desirable to better understand the physics of this object. In
particular, models are predicting a steep increase of the spectral
energy distribution at optical wavelengths. However, L dwarfs like
HR\,7672\,B are expected to be very faint in the visible
($m_I=19.59$ and $m_R=22.00$) according to evolutionary models. A
high-contrast imaging instrument will be required to study the L
dwarf sample in the visible.
\begin{figure}
\centering
\resizebox{8cm}{8cm}{\includegraphics[width=8cm]{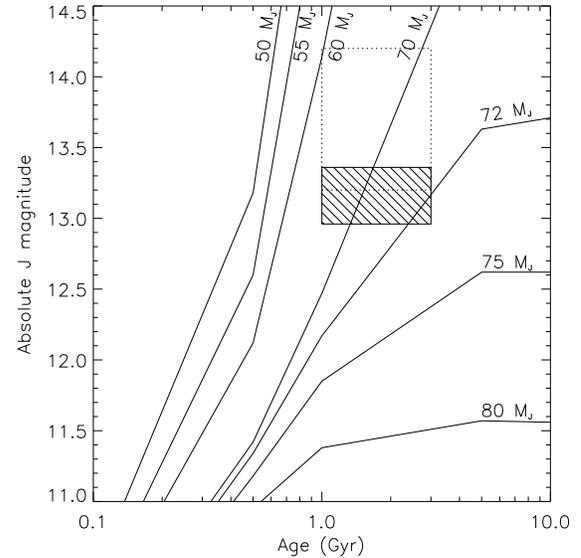}}
\caption[]{Absolute magnitude in $J$ filter as a function of the
age (in Gyr) assuming the DUSTY model of \citet{chabrier00}. The
error box for HR\,7672\,B is overplotted assuming an age of 1 to 3
Gyr. The large dash-dotted box gives the possible range of
absolute $J$ magnitude consistent with the L6$\pm$1.5 spectral
type.}
\end{figure}

\begin{acknowledgements}
The authors would like to thank the Palomar staff for its
efficient support during the observing run in July 2002, and the
referee for helpful comments to improve consistency and
readability of this paper.
\end{acknowledgements}

%%%%%%%%%%%%%%%%%%%%%%%%%%%%%%%%%%%%%%%%%%%%%%%%%%%%%%%%%%%%%%%%%%
%\begin{thebibliography}{}
\bibliography{ms3808biblio}
%\end{thebibliography}

\end{document}